\documentclass[conference,letterpaper]{IEEEtran}
\IEEEoverridecommandlockouts

\usepackage{cite}
\usepackage{svg}
\usepackage{amsmath,amssymb,amsfonts}
\usepackage{algorithmic}
\usepackage{graphicx}
\usepackage{xcolor}

\definecolor{dark_red}{HTML}{C1272D}
\definecolor{indigo}{HTML}{0000a7}
\definecolor{yellow}{HTML}{eecc16}
\definecolor{teal}{HTML}{008176}
\definecolor{light_gray}{HTML}{b3b3b3}

\usepackage{textcomp}
\usepackage{paralist}

\usepackage[utf8]{inputenc}

\usepackage{url}

\usepackage[caption=false,font=footnotesize]{subfig}

\usepackage{booktabs}

\usepackage{listings}
\makeatletter
\newcommand{\linebreakand}{%
  \end{@IEEEauthorhalign}
  \hfill\mbox{}\par
  \mbox{}\hfill\begin{@IEEEauthorhalign}
}

\usepackage{todonotes} 
\def\BibTeX{{\rm B\kern-.05em{\sc i\kern-.025em b}\kern-.08em
    T\kern-.1667em\lower.7ex\hbox{E}\kern-.125emX}}

\begin{document}


\title{User Preferences for Large Language Model versus Template-Based Explanations of Movie Recommendations: A Pilot Study}

\author{
    \IEEEauthorblockN{Julien Albert}
    \IEEEauthorblockA{UNamur\\julien.albert@unamur.be}

    \and

    \IEEEauthorblockN{Martin Balfroid}
    \IEEEauthorblockA{UNamur\\martin.balfroid@unamur.be}   

    \and 

    \IEEEauthorblockN{Miriam Doh}
    \IEEEauthorblockA{ULB-UMONS\\miriam.doh@ulb.be}

    \and

    \IEEEauthorblockN{Jeremie Bogaert}
    \IEEEauthorblockA{UCLouvain\\jeremie.bogaert@uclouvain.be}

    \linebreakand

    \IEEEauthorblockN{Luca La Fisca}
    \IEEEauthorblockA{UMONS\\luca.lafisca@umons.be}

    \and

    \IEEEauthorblockN{Liesbet De Vos}
    \IEEEauthorblockA{UNamur\\liesbet.devos@unamur.be}

    \and

    \IEEEauthorblockN{Bryan Renard}
    \IEEEauthorblockA{Multitel-UNamur\\renard@multitel.be\\bryan.renard@unamur.be}

    \and

    \IEEEauthorblockN{Vincent Stragier}
    \IEEEauthorblockA{UMONS\\vincent.stragier@umons.ac.be}

    \and

    \IEEEauthorblockN{Emmanuel Jean}
    \IEEEauthorblockA{Multitel\\jean@multitel.be}
}

\maketitle

\begin{abstract}

Recommender systems have become integral to our digital experiences, from online shopping to streaming platforms. Still, the rationale behind their suggestions often remains opaque to users. While some systems employ a graph-based approach, offering inherent explainability through paths associating recommended items and seed items, non-experts could not easily understand these explanations. A popular alternative is to convert graph-based explanations into textual ones using a template and an algorithm, which we denote here as “template-based” explanations. Yet, these can sometimes come across as impersonal or uninspiring. A novel method would be to employ large language models (LLMs) for this purpose, which we denote as “LLM-based”. To assess the effectiveness of LLMs in generating more resonant explanations, we conducted a pilot study with 25 participants. They were presented with three explanations: (1) traditional template-based, (2) LLM-based rephrasing of the template output, and (3) purely LLM-based explanations derived from the graph-based explanations. Although subject to high variance, preliminary findings suggest that LLM-based explanations may provide a richer and more engaging user experience, further aligning with user expectations. This study sheds light on the potential limitations of current explanation methods and offers promising directions for leveraging large language models to improve user satisfaction and trust in recommender systems.

\end{abstract}


\section{Introduction}

Most of us wonder daily why platforms like Facebook and YouTube recommend specific people or videos to us. The lack of transparency in these recommendations often leaves us without a clear explanation. This can degrade user confidence, recommendation acceptance and, more generally, the user experience~\cite{Tintarev2015}. To address those important concerns, a growing field of research focuses on making recommendation systems more transparent and explainable~\cite{Papadimitriou2012, Tintarev2015, Zhang2020}. A promising approach is to use large language models (LLMs) to generate explanations\footnote{In the context of this abstract, when we refer to the generation of LLMs-based recommendations explanations, we actually mean using an LLMs to rephrase an explanation or interpret a graph representation of an explainable recommendation.} for recommendations. LLMs are initially pre-trained on extensive corpora, allowing them to perform a versatile range of natural language processing (NLP) tasks\cite{bommasani2021foundationModels}. The generated text is typically well-written and clear, making it easy for humans to understand.

Motivated by these perspectives, we put them to the test in the generation of explanations for recommendations during the TRAIL’23 Workshop\footnote{\url{https://trail.ac/en/trail-summer-workshops/the-trail-summer-workshop-2023/}, more details in the Appendix}. Concretely, we defined two goals to address during the workshop.
The first is implementing working examples of recommendation explanations generated with LLMs using various recommendation methods and LLM models. This way, we could assess the technical possibilities and limitations of LLMs.
The second goal is to evaluate explanations generated by different LLM models and recommendation methods to understand their qualities and their limitations in this context. To achieve this goal, we designed a user-based evaluation method to assess explanations w.r.t. different explanatory goals and subjective properties~\cite{Tintarev2015}.

\section{Technical Exploration and Implementation}

\begin{figure}[!ht]
    \centering
    \includegraphics[width=\linewidth]{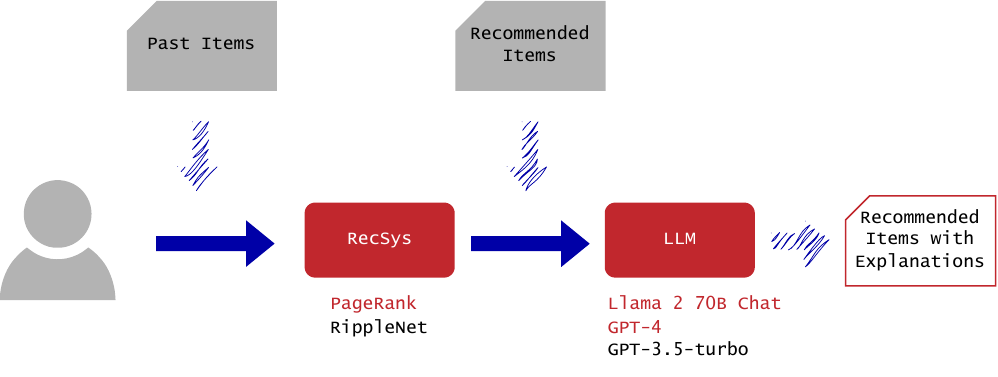}
    \caption{Pipeline used to guide our experiments. Methods and models used for the evaluation part are in fuchsia.}
    \label{fig:Pipeline}
\end{figure}

As shown in Fig.~\ref{fig:Pipeline}, we propose a pipeline that takes user preferences (i.e., past interactions with items) as input, and generates explained recommendations as output.
The most important design choice is to separate the recommendation and explanation processes, only using LLMs to explain items previously recommended by a standalone recommendation method.
We choose to use classic recommendation to ensure valid recommendation, as hallucination is an important issue with LLMs~\cite{ji2023hallucination}. Moreover, this choice allows us to isolate the explanation task, empowering us to compare explanations created by a baseline explanation method, with explanations written by LLMs.

Regarding the recommendation methods, we focused on graph-based methods. More specifically, we used \textit{Personalized PageRank}~\cite{haveliwala2003personalizedpagerank} and RippleNet~\cite{wang2018ripplenet}, both of which generate explanations based on a graph of the past interactions between users and items. This graph is augmented with knowledge about the movie domain, to further guide the recommendation system.
Those methods also provide explanations for recommendations in the form of paths from the seed items to the recommended ones. The datasets used for experimentation were Movielens-1M\footnote{\url{https://grouplens.org/datasets/movielens/}} and MindReader\footnote{\url{https://mindreader.tech/dataset/}}. For the user-based evaluation, we only used Movielens-1M in combination with the \textit{Personalized PageRank}.

We are interested in textual explanations, since they convey rich information to the user~\cite{Tintarev2015}. The main existing approaches are template-based and generation-based~\cite{Zhang2020}. As a baseline, we use a template-based approach, which transcripts path-based explanations into text. We compare this baseline method to LLM-based methods for generating explanations, inspired by the literature on the topic, e.g., PEPLER~\cite{Li2023PersonalizedPromptLearningExplainableRecommendation}.


Large Language Models (LLMs) are now some of the world's most famous NLP models due to the publicity made by OpenAI with ChatGPT, which uses GPT-3.5-turbo and GPT-4 (SOTA). They can perform various NLP tasks. Current LLMs use a decoder-only architecture based on the transformer's architecture\cite{yangHarnessingPowerLLMs2023a}. They are trained to give a probability of distribution over the vocabulary of tokens, allowing to predict the next token.
The tokens are subparts of sentences, and the vocabulary of tokens, fixed and based on the training data, is often built using byte pair encoding (BPE)\cite{radfordImprovingLanguageUnderstanding, touvronLlamaOpenFoundation2023}. To produce sequences of tokens, we used greedy decoding with Llama 2 70B Chat and the default technique (which we don't know of) when using GPT-4. Greedy decoding only considers the most probable token at each generation step, which is time-efficient, unlike other techniques. We decided on using greedy decoding for our exploration study but plan to investigate other strategies in the future.


We considered two methods for generating explanations for movie recommendations. We aimed to measure how effectively each approach could deliver concise yet informative explanations to users that align with their expectations.

\begin{figure}[!ht]
    \centering
    \includegraphics[width=0.6\linewidth]{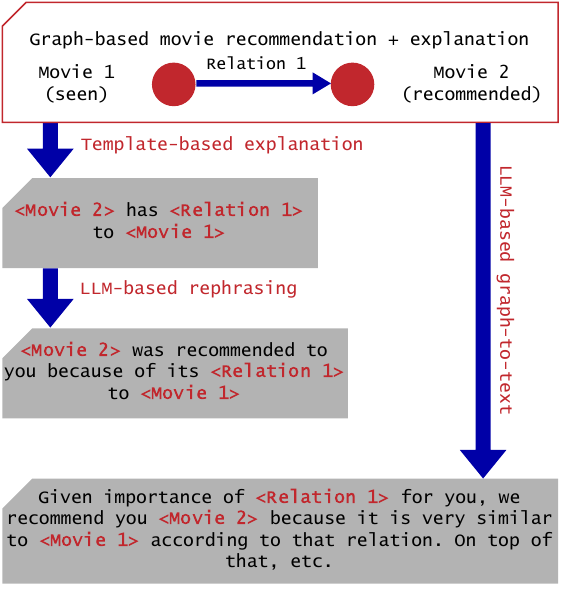}
    \caption{Illustration of the three types of explanations compared in the user evaluation.}
    \label{fig:Example}
\end{figure}

Three types of explanations were finally chosen for the user-based evaluation (as shown in Fig.~\ref{fig:Example}):
\begin{enumerate}
    \item \textbf{Template-based}: our baseline method, which uses a template to generate explanations algorithmically based on the edges and nodes of the explanation paths;
    \item \textbf{LLM-based}: which uses LLMs to generate the explanation. We explored two variations:
          \begin{enumerate}
              \item \textbf{LLM-based rephrasing}: rephrase the template-based explanation;
              \item \textbf{LLM-based graph-to-text}: the model deduces the reasoning behind the recommendation given a knowledge graph as context.
          \end{enumerate}
\end{enumerate}

Between the two LLM variants, only the context varies, either the template-based explanation or the graph. The definition of the task is, therefore, the same for both: to explain why a particular film has been recommended. To ensure a fairly consistent format across each generation, we constrained the LLM's behaviour~\cite{reynolds2021prompt} by specifying that only one paragraph should be used and that it should be written in layman's terms. Otherwise, the model tended to ramble and use technical terms that could confuse the user.

\begin{figure}[!ht]
    \centering
    \subfloat[LLM-based rephrasing\label{fig:llm-based-rephrasing-prompt}]{%
        \includegraphics[width=0.45\textwidth]{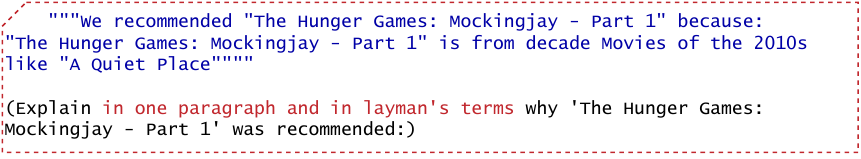}}
    \\
    \subfloat[LLM-based graph-to-text\label{fig:llm-based-graph2text-prompt}]{%
        \includegraphics[width=0.45\textwidth]{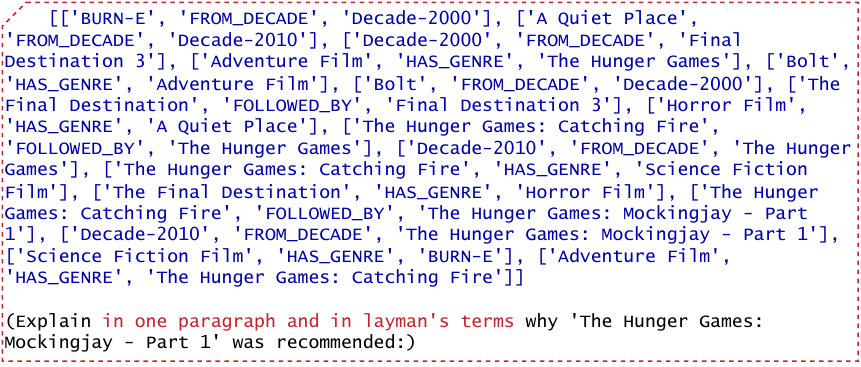}}

    \caption{
        {Here is an example of the same recommendation presented in the same format as the prompt in Liu et al.~\cite{liu2023chatgpt}. \textbf{Black}-colored text outlines the task, \color{dark_red}\textbf{red}}-colored text highlights the formatting guidelines, and
            {\color{indigo}\textbf{blue}}-colored text is either the given template or the graph.
    }
\end{figure}

\section{User-based evaluation} 

\subsection{Methodology}

\begin{figure}[!ht]
    \centering
    \includegraphics[width=\linewidth]{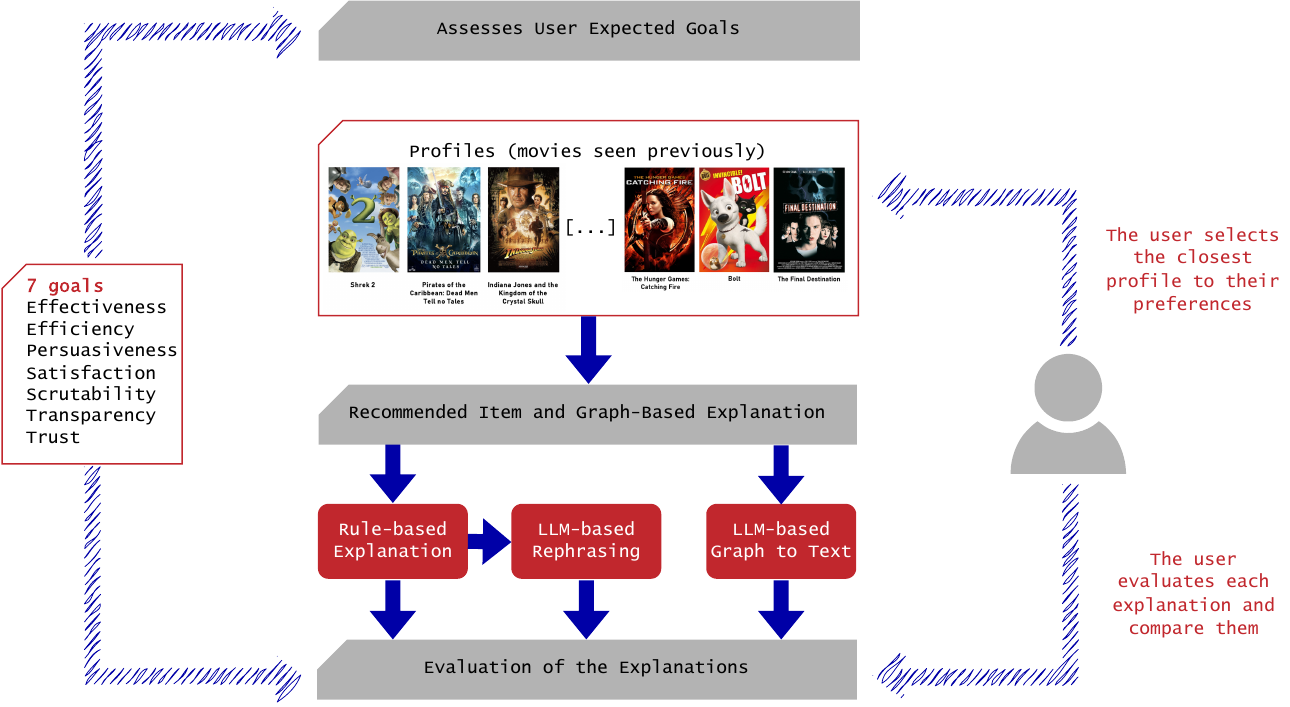}
    \caption{Structure of our user evaluation procedure}
    \label{fig:UserEvalStructure}
\end{figure}

Part of our project's goal was to perform a user-based evaluation of the three types of explanations generated by our pipeline. We drew inspiration from \cite{balog2020measuring} to craft the structure for our evaluation procedure (Fig.~\ref{fig:UserEvalStructure}), albeit with slight modifications due to the inclusion of LLM-generated explanations. We decided to focus on the following key aspects:
\begin{enumerate}
    \item Assessing user expectations of recommendation explanations using the seven goals from \cite{Tintarev2015}, also used by \cite{balog2020measuring}.
    \item Presenting a recommended item to the user alongside multiple alternative explanations (based on a watching profile selected by the user beforehand).
    \item Requesting users to assess the explanations based on their general preference and measure the extent to which each explanation satisfies the seven goals.
    \item Gathering qualitative insights via open question on user expectations and explanation assessments.
\end{enumerate}

Afterward, we conducted a dry run to validate the clarity of the questionnaire and assess its length to prevent evaluator fatigue. Subsequently, we rolled out the questionnaire to multiple evaluators to gather their responses.

\subsection{Results}

We conducted 25 user tests with TRAIL'23 Workshop participants (researchers in AI). The small number of participants means that no statistically robust conclusions can be drawn, but certain trends can be observed.

\begin{figure}[!ht]
    \centering
    \includegraphics[width=\linewidth]{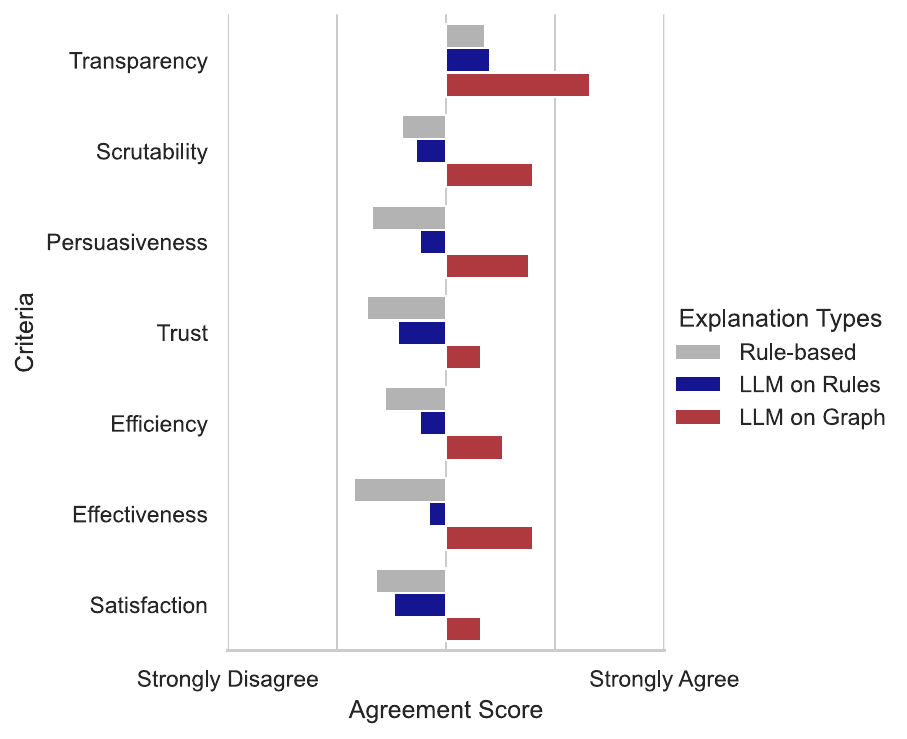}
    \caption{User assessment of explanations w.r.t. the 7 goals. about the recommendation explanations.}
    \label{fig:ResultsExplanation}
\end{figure}

Concerning the user's expectations about explanations, we observe no difference in importance for the seven goals investigated. However, concerning user assessment of the generated explanations (see Fig.~\ref{fig:ResultsExplanation}), we observe that the explanation generated by the LLM from a knowledge graph performs best w.r.t. the 7 goals. And this result is confirmed by the participants' general assessment of the explanations. According to the participants, this explanation type is mainly preferred because it's often more detailed and more pleasant to read.
However, beyond the small sample size ($n = 25$), it is important to point out the significant variance in these last results. This indicates strong differences between participants in the way they perceive explanations, which is a result that should be investigated further.

Moreover, we observed that LLMs often introduce additional details. Although most of them seem correct, they do not come from the knowledge graph and are therefore not verified. This may be due to the model's ability to draw on cultural references~\cite{reynolds2021prompt}, in this case, movie titles. If this is undesirable, a workaround is to use numbered labels instead of movie titles. This will limit the model only to use in-context information.

\section{Future Works}

In the future, we would like to explore various models. In particular, we would like to focus on smaller LLMs, to understand how model size affects efficiency. We expect that, due to their size, smaller models may struggle to generate explanations based on the knowledge graph, but may be sufficient for rephrasing the template-based explanations. This is particularly relevant considering the substantial computational requirements of larger LLMs. Furthermore, fine-tuning could be explored, as well as advanced prompting techniques like chain-of-thought~\cite{weichain} and self-consistency~\cite{wang2022self} (which may appeal to users wanting more detailed reasoning). Another interesting approach might be to specify the explanation generation task by memetic proxy~\cite{reynolds2021prompt}, i.e., to use the model's ability to draw on cultural references, metaphors, analogies, role-playing, and so on.

Finally, instead of only relying on user-based evaluations, we aim to use mixed-methods evaluation to draw a complete picture of LLM's explanation generation capabilities for recommendations. This evaluation would combine heuristics-based methods (based on classical metrics for text quality like BLEU~\cite{papineni2002bleu} and ROUGE~\cite{lin2004rouge} scores), explanation quality metrics (e.g., \cite{Li2023PersonalizedPromptLearningExplainableRecommendation}), and user-based methods. Such user-based methods could include qualitative (e.g., interviews) and quantitative (e.g., online survey) methods to assess explanations w.r.t. different explanatory goals and subjective properties~\cite{Tintarev2015}.

\section*{Acknowledgment}
This research was partially supported by the ARIAC project (No. 2010235), funded by the Service Public de Wallonie (SPW Recherche).
This research used resources of the “Plateforme Technologique de Calcul Intensif (PTCI)” (http://www.ptci.unamur.be) located at the University of Namur, Belgium, which is supported  by the FNRS-FRFC, the Walloon Region, and the University of Namur (Conventions No. 2.5020.11, GEQ U.G006.15, 1610468, RW/GEQ2016 et U.G011.22). The PTCI is member of the “Consortium des Équipements de Calcul Intensif  (CÉCI)” (\url{https://www.ceci-hpc.be}).
Vincent Stragier is funded through a PhD grant from the Œuvre fédérale Les Amis des Aveugles et Malvoyants ASBL- The Friends of the Blind and Visually Impaired Federal Charity-, Ghlin, Belgium and the Loterie Nationale, Rue Belliard 25-33, 1040 Brussels, Belgium. Vincent Stragier is partially supported by the FNRS-FRS.
Bryan Renard is funded by the Public Service of Wallonia (Economy, Employment and Research), under the FoodWal agreement n°2210182 from the Win4Excellence project of the Wallonia Recovery Plan.

\bibliographystyle{ieeetr}
\bibliography{references}

\newpage
\appendix
\subsection{Authors' Biographies}


\subsubsection{Julien Albert}

After an initial career as a librarian, Julien Albert embarked on a career change and obtained a master's degree in computer science from UNamur in 2020. He then worked for one year at UNamur on the EFFaTA-MeM research project, which aims to develop innovative tools for text analysis. In September 2021, he began a Ph.D. in computer science at UNamur under the supervision of Professors Benoît Frenay and Bruno Dumas. His research area is explainability in artificial intelligence. His approach involves placing the user at the centre of concerns by combining explainability techniques from machine learning with methods developed in human-computer interaction.

\subsubsection{Martin Balfroid}

Martin Balfroid is a PhD student at the University of Namur, his research investigates AI-in-the-loop approaches to improve software engineering. He earned his master's in Computer Science, focusing on Data Science, in June 2022. The results of his master's thesis were published at the 2nd Software Testing Education Workshop. Martin began his PhD in July 2022 with funding from the ARIAC project and is supervised by Assistant Professors Benoît Vanderose and Xavier Devroey. 

\subsubsection{Miriam Doh}

Miriam Doh obtained a master's degree in Information and Communication Engineering from the University of Trento (UniTn) in Italy in 2021. Her master's thesis focused on the application of genetic algorithms to social networks, with a focus on studying the problem of community segregation in metropolitan areas. After completing her degree, she began a joint PhD program between ULB and UMONS on the intersection of Deep Learning and Computer Vision, with a particular emphasis on Explainable AI (XAI). Her research project is dedicated to exploring the integration of Cognitive Psychology principles to advance Explainable and Trustworthy Artificial Intelligence, particularly within the context of Face Analysis applications.

\subsubsection{Jeremie Bogaert}

Jérémie Bogaert obtained his master's degree in computer science engineering with a focus on artificial intelligence from UCLouvain in 2021. His master's thesis explored the limitations of deep fake news generation models and their detection using machine learning models and human readers. He started his doctoral thesis at UCLouvain in September 2021 and is currently working on the interaction between interpretable machine learning models and human readers for the detection of deep fake news. He has interest in studying the explainability of NLP models in general.

\subsubsection{Luca La Fisca}

Luca La Fisca is currently a PhD student with a keen interest in neural engineering. His primary research focus revolves around advancing tools for a deeper understanding of the intricacies of the human brain. Luca's doctoral thesis specifically delves into the realm of ElectroEncephalogram (EEG) analysis. He is particularly fascinated by the interpretation of latent space to unveil critical interactions among brain regions during the execution of specific tasks, with a primary emphasis on visual tasks. Additionally, Luca harbours a strong interest in the field of neurofeedback.

Within the ARIAC project, Luca La Fisca is actively involved in Work Package 1, which centres on the interactions between humans and artificial intelligence. His contributions span various aspects, including interactive and human-in-the-loop algorithms, user assistance in AI-in-the-loop scenarios, consensus mechanisms, handling imperfect multi-expert labels, and the development of explainable AI solutions.

\subsubsection{Liesbet De Vos}

Liesbet De Vos obtained a master's in Linguistics at the Catholic University of Leuven in 2021. Fascinated by computational linguistics, she completed her studies with an advanced master's in Artificial Intelligence at the Catholic University of Leuven, which she completed in 2022. Liesbet continues to nurture her passion for language during a PhD at the University of Namur, where she focuses on building hybrid AI systems that learn to use language through the same mechanisms as humans. In her thesis, she aims to extend the computational construction grammar framework to the visual modality so that it can adequately represent and learn the linguistic structure of sign languages. Within the ARIAC project, Liesbet actively contributes to Work Package 2, which revolves around trust mechanisms for artificial intelligence. 

\subsubsection{Bryan Renard}

Bryan Renard obtained a master's degree in theoretical physics from UNamur in 2022. He then changed his career path and is now a dedicated PhD student whose research interests span several exciting domains within the field of artificial intelligence. His primary focus is the application of artificial intelligence in the realm of proteins, exploring innovative ways to harness AI (especially LLMs) for protein-related research. Additionally, Bryan is passionate about self-supervised learning, particularly in the context of Automatic Speech Recognition (ASR).

His thesis is jointly conducted by UNamur and Multitel. It is funded by the FoodWal portfolio from the Public Service of Wallonia (Economy, Employment, and Research), more particularly within the PEPTIBoost project. As a part of the ARIAC project, Bryan Renard plays an integral role in Work Package 4, which revolves around optimizing AI implementations. His contributions encompass a wide range of topics, including transfer learning, High-Performance Computing (HPC) and self-supervised learning techniques.

\subsubsection{Vincent Stragier}

Vincent Stragier is a PhD student at the University of Mons (UMONS). He is working on an interactive assistant for visually impaired and blind people within the ISIA Lab, a department of the Faculty of Engineering. His research interests are mainly focused on NLP, large language models and computer vision related topics.

In 2021, he obtains his master’s degree in electrical Engineering, specialized in Signals, Systems and BioEngineering from the Faculty of Engineering in Mons. In 2020, he works on an epilepsy detection pipeline base on an XGBoost classifier built by the CETIC, where he is Engineer Intern at the time. During his studies, he participates in the electronic student association, electroLAB, and the Erasmus Student Network of Mons, ESNMons. In his free time, he likes taking photographs, fixing various things (hardware and software related), and learning new skills.

\subsubsection{Emmanuel Jean}

In 2009, Emmanuel Jean earned a dual degree in electrical engineering from the Faculty of Engineering at the University of Mons and Supelec-Paris. Subsequently, he joined the Signal Processing and Embedded Systems department at Multitel, where he actively participated in various regional and European projects involving vocal technologies and multimodal Human-Computer Interaction (HCI).

In 2012, he furthered his education by obtaining a Bachelor's degree in Management Sciences from the Louvain School of Management at UCL-Mons. Since 2017, his professional focus has shifted towards diverse projects centred around Deep Learning applied to temporal signals, including audio, speech, and vibrations. His current research interests revolve around the development of weakly supervised machine learning techniques and the deployment of reliable artificial intelligence systems.


\subsection{ARIAC and TRAIL}

TRAIL and the ARIAC research project are part of the regional DigitalWallonia4.ai program, which aims to accelerate the development of artificial intelligence technologies in Wallonia.

TRAIL (TRusted AI Labs) provides actors in the socio-economic landscape with R\&D expertise and AI technological bricks developed by the 5 French-speaking universities and 4 approved research centres active in AI. To achieve this, the SPW-EER has allocated a budget of €32 million for the ARIAC research project led by the TRAIL consortium. This initiative is part of the 4th axis of the regional DigitalWallonia4.ai programme: “Research, innovation and partnerships”.

The ambition is to pool research in artificial intelligence in the Wallonia-Brussels Federation and is concretely reflected through the research project “ARIAC by DigitalWallonia4.ai”, based on an agreement between the Walloon Region (SPW Research) and the actors forming the TRAIL consortium.

The ARIAC project (“Applications and Research for Trusted Artificial Intelligence” in English or “Applications et Recherche pour une Intelligence Artificielle de Confiance” in French) is spread over 6 years and is articulated around 5 WP (Work Package):

\begin{itemize}
    \item human-AI interaction,
    \item trust mechanisms for AI,
    \item model-AI integration,
    \item optimized implementations of AI,
    \item TRAIL Factory.
\end{itemize}

\end{document}